\documentstyle[12pt]{article}
\input psfig.tex
\begin{document}

\def\beq{\begin{equation}}
\def\eeq{\end{equation}}
\def\D0{D\0~}
\def\ov{\overline}
\def\ra{\rightarrow}
\def\pslash{\not{\hbox{\kern -1.5pt $p$}}}
\def\kslash{\not{\hbox{\kern -1.5pt $k$}}}
\def\aslash{\not{\hbox{\kern -1.5pt $a$}}}
\def\bslash{\not{\hbox{\kern -1.5pt $b$}}}
\def\Dslash{\not{\hbox{\kern -4pt $D$}}}
\def\wslash{\not{\hbox{\kern -4pt $\cal W$}}}
\def\zslash{\not{\hbox{\kern -4pt $\cal Z$}}}
\def\ttz{{\mbox {\,$t$-${t}$-$Z$}\,}}
\def\tta{{\mbox {\,$t$-${t}$-$A$}\,}}
\def\tbw{{\mbox {\,$W$-${t}$-$b$}\,}}
\def\tcZ{{\mbox {\,$t$-${c}$-$Z$}\,}}
\def\tuZ{{\mbox {\,$t$-${u}$-$Z$}\,}}
\def\tcg{{\mbox {\,$t$-${c}$-$g$}\,}}
\def\Tr{{\rm Tr}}
\def\fb{${\rm fb}^{-1}$}
\def\tsmdecay{$t \ra b \; W \ra b \; \ell \; \bar{\nu_{\ell}} \;$}

\setcounter{footnote}{1}
\renewcommand{\thefootnote}{\fnsymbol{footnote}}

\begin{titlepage}

{\small
\noindent
{hep-ph/9710372} \hfill {MSUHEP-71015} \\
{ October 1997} \hfill {ANL-HEP-PR-97-85} \\

}

\vspace{2.0cm}

\begin{center}
{\Large\bf The Phenomenology of Single Top Quark
Production at the Fermilab Tevatron}
\end{center}

\vspace*{1.2cm}
\baselineskip=17pt
\centerline{\normalsize  
Tim Tait\footnote{
\baselineskip=12pt  Also at Argonne National Laboratory,
HEP Division, 9700 South Cass Avenue,  Argonne, IL 60439.},
  and C.-P. Yuan}
 
\centerline{\normalsize\it
Department of Physics and Astronomy, Michigan State University }
\centerline{\normalsize\it
East Lansing, Michigan 48824 , USA }

\vspace{0.4cm}
\raggedbottom
\setcounter{page}{1}
\relax

\begin{abstract}
\noindent
Single top quark production at the Fermilab
Tevatron Run II (a $p \bar{p}$ collider
with $\sqrt{S} = 2$ TeV) proceeds dominantly
via two sub-processes, a
$t$-channel $W$-gluon fusion process, and an $s$-channel
$W^*$ process.  We show that these two
sub-processes have different sensitivities to new physics effects
in the top quark's electro-weak interactions.  The $W^*$ process
is sensitive to new heavy charged resonances, such as a $W'$
boson, while the $W$-gluon fusion process is more sensitive to
modifications to the top's interaction, including
flavor-changing neutral currents involving the top
quark.
We examine the implications of these results on our ability
to measure $V_{tb}$ with confidence, and propose a quantity
$R = \sigma_{Wg} / \sigma_{W^*}$,
which may be studied in order to characterize the confidence
one may place upon a given measurement of $V_{tb}$ from
single top production.
\end{abstract}

\vspace*{3.4cm}
\end{titlepage}

\normalsize\baselineskip=15pt
\setcounter{footnote}{0}
\renewcommand{\thefootnote}{\arabic{footnote}}

\section{Introduction}
\label{intro}
\indent \indent

With the discovery of the top quark by the CDF and
D$\emptyset \;$
collaborations \cite{topdisc}, it has become natural to
consider its properties, such as
its couplings to the other particles
of the Standard Model (SM).
The top
quark is singled out by its large mass
($m_t \simeq$ 175 GeV), the
same order as the electro-weak symmetry breaking (EWSB)
scale, $v = 246$ GeV.  Thus it seems reasonable that the
top may prove valuable in probing the mechanism of mass
generation in the Standard Model of particle physics.
Indeed, if there is some type of new physics associated with
the generation of mass, it may be more apparent in the top
quark sector than for any of the other, lighter, known fermions,
in accordance with the mass hierarchy.

While production of $t \bar{t}$ pairs \cite{ttbar}
provides an excellent opportunity to probe the top's
QCD properties,
in order to carefully measure the top's electro-weak interactions
it is also useful to consider single top production, in
addition to studying the decay of the top quark in $t \bar t$
events.
Single top production at the Tevatron
occurs within the SM in three
different channels, the $s$-channel $W^*$ production,
$q q' \ra W^* \ra t \bar{b}$
\cite{schan, schan1, schan2, schanresum, boos},
the
$t$-channel $W$-exchange mode, $b q \ra t q'$
\cite{boos, tchan, tchan1, tchan2, cpy, dougthesis, bordes, tchan3}
(sometimes referred to as $W$-gluon
fusion), and through $t W^{-}$ production \cite{tw}.
These three sub-processes have very different kinematics
and experimental signatures \cite{schan1, tchan1, cpy, dougthesis},
and as we shall discuss later, are sensitive to different types
of new physics in the top quark sector.  Thus they
provide complimentary information about
the properties of the top quark.

In this paper, we quantify the observation that the modes
of single top production can be used to provide complimentary
information about the top quark.  In Section \ref{singlet}, we
discuss the properties of these separate modes of
single top production in the context of the SM,
summarizing the positive and negative aspects of each for
probing the top quark sector at the Tevatron
Run II\footnote{The Fermilab Tevatron is a $p \bar{p}$
collider.  At Run II it is expected that a center-of-mass energy
$\sqrt{S} = 2$ TeV for the colliding protons and anti-protons
will be realized.  For the purposes of this study,
we consider a data sample with an integrated luminosity of
$L = 2, 10$, and 30 \fb.}.  In
Section~\ref{resonance} we show the effect of an additional heavy
resonance on each sub-process, and in Section \ref{width} we
examine the effects from possible modifications to the top's
couplings to the other particles of the SM.  Finally, we
examine what the possibility of new physics in the top quark
sector implies about our ability to measure $V_{tb}$ with
confidence.

\section{Single Top Production
at a Hadron Collider}
\label{singlet}
\indent\indent

\begin{figure}
\centerline{\hbox{
\psfig{figure=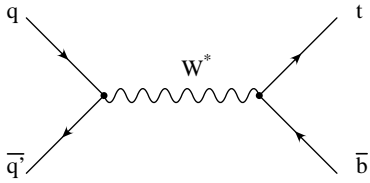,height=1.0in}}}
\caption{Feynman diagram for
$q \bar{q} \ra W^* \ra t \bar{b}$ at the leading order.}
\label{schannelfig}
\end{figure}

\begin{figure}
\centerline{\hbox{
\psfig{figure=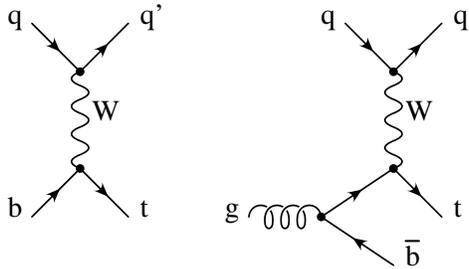,height=1.5in}}}
\caption{ Representative Feynman diagrams for
the $W$-gluon fusion mode of single-top production.
The gluon splitting diagram is not really an $\alpha_S$
QCD correction, but rather a 1/${\rm ln}(m_t^2/m_b^2)$
correction coming from the definition of the $b$ PDF.
When combining the contributions from these two diagrams,
it is necessary to subtract the part in the gluon splitting
diagram where the gluon becomes collinear with the $b$ parton,
to avoid double counting this region of kinematics.}
\label{tchannelfig}
\end{figure}

\begin{figure}
\centerline{\hbox{
\psfig{figure=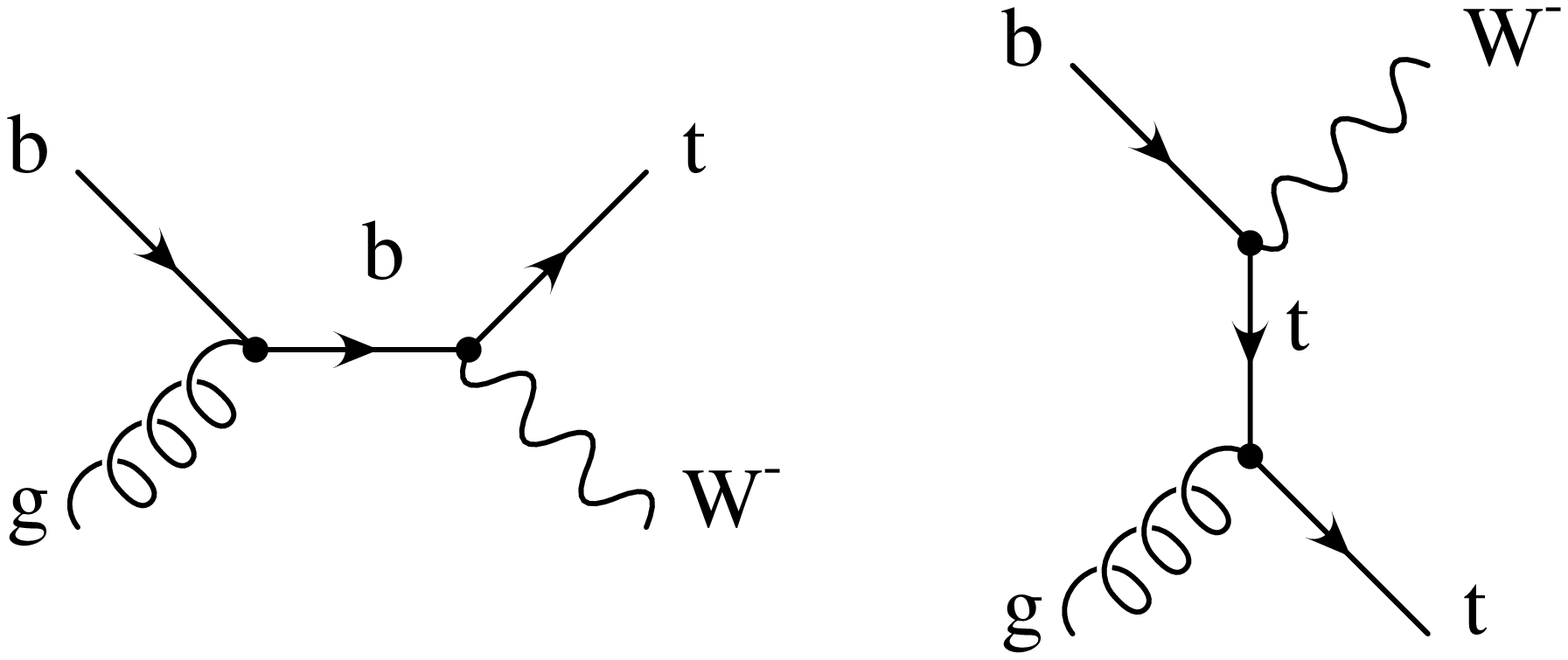,height=1.3in}}}
\caption{Feynman diagrams for
$g b \ra t W^{-}$.}
\label{twfig}
\end{figure}

Single top production at a hadron collider occurs dominantly
through three sub-processes.  The $W^*$ mode of production
shown in Figure \ref{schannelfig} occurs when a quark and an
anti-quark fuse into a virtual $W$ boson, which then splits
into a $t$ and $\bar b$ quark.  The $W$-gluon fusion mode
shown in Figure \ref{tchannelfig}, occurs
when a $b$ quark fuses with a $W^{+}$ boson, producing a
top quark.
The $t W^-$ mode
occurs when a $b$ quark radiates a $W^-$,
and is shown in Figure \ref{twfig}.
This mode may be important at the Large Hadron
Collider\footnote{The LHC is a $p p$ collider with $\sqrt{S} =$
14 TeV.}
(LHC), but is highly suppressed at the Tevatron because
of the massive $W$ and $t$ particles in the final
state.  Because of its low cross section of about 0.1
pb (for $m_t = 175$ GeV)
at the Tevatron Run II \cite{cpy, tw}, and
relative insensitivity to the new physics effects we
will be considering in this work, we do not present
detailed studies of its cross section.

The three single top production processes contain the \tbw
vertex of the SM, and thus are sensitive to any possible
modification of this vertex from physics beyond the SM,
and to the Cabibbo-Kobayashi-Maskawa
(CKM) parameter $V_{tb}$.  It has been shown
\cite{schan1, tchan1, cpy, dougthesis} 
that because of the difference in
kinematics, it is possible to statistically disentangle the
two sub-processes from each other, and from the SM backgrounds
expected at the Tevatron Run II.

\begin{figure}
\centerline{\hbox{
\psfig{figure=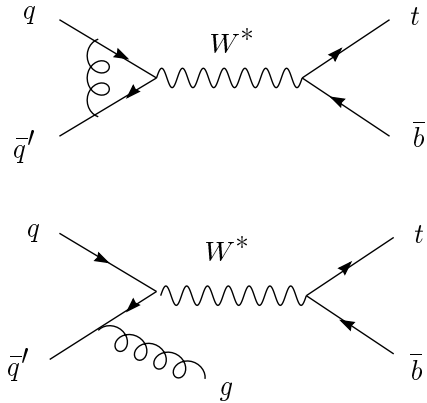,height=3.0in} }}
\caption{ Representative Feynman diagrams for NLO corrections to
$q \bar{q} \ra W^* \ra t \bar{b}$ arising from initial state
corrections.}
\label{NLOsi}
\end{figure}

\begin{figure}
\centerline{\hbox{
\psfig{figure=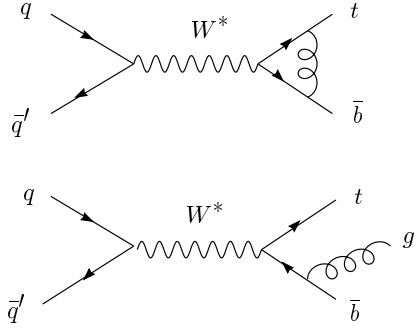,height=2.2in}}}
\caption{ Representative Feynman diagrams for NLO corrections to
$q \bar{q} \ra W^* \ra t \bar{b}$ arising from final state
corrections.}
\label{NLOsf}
\end{figure}

\begin{figure}
\centerline{\hbox{
\psfig{figure=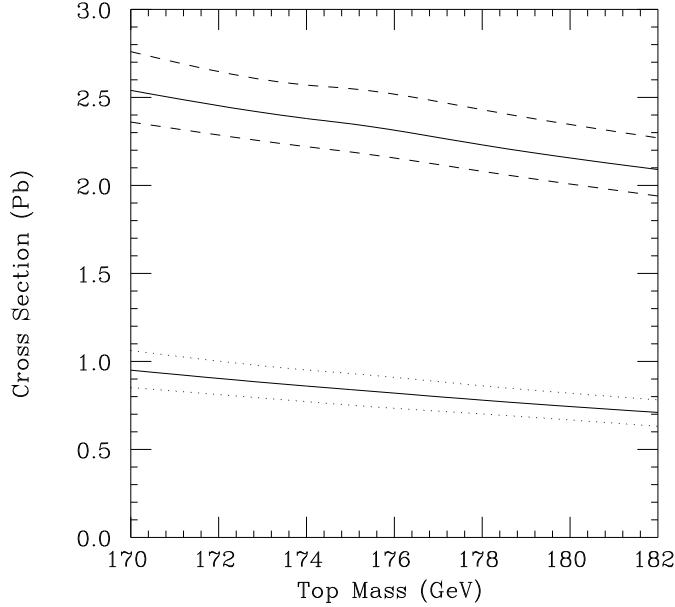,height=3.0in}}}
\caption{ The dependence of the mean values of the NLO total
cross sections, $\sigma_{Wg}$ (upper curve)
and $\sigma_{W^*}$ (lower curve),
on the mass of the top quark.  The dashed lines represent
the upper and lower bounds on $\sigma_{Wg}$, and the dotted
lines represent the upper and lower bounds on $\sigma_{W^*}$.}
\label{topmassdep}
\end{figure}

The total cross section  for the $W^*$ production sub-process
($\sigma_{W^*}$) has been studied at next-to-leading-order (NLO) in QCD
corrections (order $\alpha_{EW}^2 \alpha_S^1$) \cite{schan2}, by
including NLO corrections to both the $q \bar q'$ initial state (c.f.
Figure \ref{NLOsi}) and the $t \bar b \;$ final state (c.f. Figure
\ref{NLOsf}). Detailed studies for the kinematics of this process at
leading order (LO) \cite{schan1, boos, dougthesis} exist in the
literature, as well as  more realistic studies in which the effects of
soft gluons on the kinematics have been re-summed \cite{schanresum}.
This sub-process is an attractive mode for studying top quark
properties, because the initial state partons are quarks with relatively
large momentum fraction, $x$, and thus the parton densities are well
understood.  On the other hand, this mode suffers from a smaller cross
section than the $W$-gluon fusion mode, and a larger percentage
dependence on the mass of the top quark
\cite{schanresum}. In order to estimate the
uncertainty in the NLO theoretical prediction for the $W^*$ production
rate, we examine the dependence of the result on the
factorization and renormalization scales (which in principle are
separate quantities, however we will work with $\mu_{R} = \mu_{F} =
\mu$), and the dependence on the choice of the parton distribution
functions (PDF) used. A physical observable cannot depend on the scale
$\mu$, and a theoretical calculation to all orders in perturbation
theory must similarly be independent of the value of $\mu$ used in the
calculation.  However, at fixed order in the perturbation series the
theoretical prediction will still show some residual dependence on
$\mu$.  Thus, by varying $\mu$, one obtains an estimate of the
dependence of the calculation on the uncalculated, higher order terms in
the perturbation series.  In \cite{schan2, schanresum} it was found that
in order to cancel potentially large logarithms in the NLO perturbation
coefficients, a scale choice of $\mu = \sqrt{\hat s}$ is appropriate,
where $\hat s$ is the center of mass energy of the incoming partons.
Thus we vary the scale between $\mu = 2 \sqrt{\hat s}$ and $\mu =
\sqrt{\hat s} / 2$ in order to estimate the scale dependence of the
calculation.  To examine the PDF dependence of the result, we choose the
CTEQ4M \cite{cteq4} and MRRS(R1) \cite{mrsa} PDF. For each of various
values of $m_t$, we combine all of these predictions (due to varied PDF
and scale) to determine the upper and lower bounds on $\sigma_{W^*}$ for
each top mass under consideration, by taking the combination of PDF and
scale choice which gives the highest and lowest result respectively.  In
this way we can study the correlated effect obtained by varying the PDF
and scale (and $m_t$). We also present the central value of
$\sigma_{W^*}$, obtained as the point midway between the upper and lower
bounds, and the mean, which is the average of the two PDF sets with the
scale choice $\mu = \sqrt{\hat s}$. The mean value represents the best
theoretical estimate of $\sigma_{W^*}$.  The fact that the mean and
central values are the same for most of the top masses considered shows
that the result does not depend more strongly on either raising or
lowering the scale, and thus indicates that the choice of scale is
appropriate. Our results are presented in Table \ref{schantab} and are
shown graphically in Figure \ref{topmassdep}.  These results include the
full CKM elements for the light quark fusion into  $W^*$, but do not
include any CKM parameters in the $W$-$t$-$b$ vertex\footnote{This is
equivalent to studying $t \bar b$ production and not $t \bar s$ or $t
\bar d$ production, under the assumption that $V_{tb} = 1$.}.  The
numbers include the rate of both $t \bar b$ and $\bar t b$ production 
through the $W^*$ mode.  If we define the variable,  
\beq  
\Delta m_t \equiv m_t - 175 \; {\rm GeV} \; ,  
\eeq  
we can parameterize the curves in
Table \ref{schantab} (in units of pb) as follows, 
\begin{eqnarray} 
\sigma^{\rm mean}_{W^*}(\Delta m_t) = 0.84 - 
(2.0 \times 10^{-2}) {\Delta m_t} +
(8.9 \times 10^{-6}) {\Delta m_t}^2 - \nonumber \\
(3.6 \times 10^{-5}) {\Delta m_t}^3 +
(8.9 \times 10^{-6}) {\Delta m_t}^4 \nonumber \\
\nonumber \\
\sigma^{\rm central}_{W^*}(\Delta m_t) = 0.84 - 
(2.0 \times 10^{-2}) {\Delta m_t} +  
(6.6 \times 10^{-5}) {\Delta m_t}^2 - \\
(8.3 \times 10^{-5}) {\Delta m_t}^3 +
(1.4 \times 10^{-5}) {\Delta m_t}^4 \nonumber \\
\nonumber \\
\sigma^{\rm upper}_{W^*}(\Delta m_t) = 0.93 - 
(2.1 \times 10^{-2}) {\Delta m_t} +  
(6.6 \times 10^{-5}) {\Delta m_t}^2 - \nonumber \\
(8.3 \times 10^{-5}) {\Delta m_t}^3 +
(1.4 \times 10^{-5}) {\Delta m_t}^4 \nonumber  \\
\nonumber \\
\sigma^{\rm lower}_{W^*}(\Delta m_t) = 0.75 - 
(1.9 \times 10^{-2}) {\Delta m_t} +  
(8.7 \times 10^{-4}) {\Delta m_t}^2 +  \nonumber \\
(4.8 \times 10^{-5}) {\Delta m_t}^3 -
(1.9 \times 10^{-5}) {\Delta m_t}^4 \nonumber \;.
\end{eqnarray}
         
\begin{table}[htbp] 
\begin{center} 
\vskip -0.06in
\begin{tabular}{|l||c|c|c|c|}\hline \hline 
$m_{t}$ (GeV) & $\sigma_{W^*}$ (mean) (pb) & 
$\sigma_{W^*}$ (central) (pb) &
$\sigma_{W^*}$ (upper) (pb) & $\sigma_{W^*}$ (lower) (pb) \\ \hline
170 & 0.95 & 0.96 & 1.06 & 0.85 \\
171 & 0.92 & 0.93 & 1.02 & 0.83 \\     
172 & 0.90 & 0.90 & 0.99 & 0.81 \\ 
173 & 0.88 & 0.88 & 0.97 & 0.79 \\ 
174 & 0.86 & 0.86 & 0.95 & 0.77 \\ 
175 & 0.84 & 0.84 & 0.93 & 0.75 \\
176 & 0.82 & 0.82 & 0.91 & 0.73 \\ 
177 & 0.80 & 0.80 & 0.89 & 0.72 \\ 
178 & 0.78 & 0.78 & 0.86 & 0.70 \\ 
179 & 0.76 & 0.76 & 0.83 & 0.69 \\ 
180 & 0.74 & 0.74 & 0.81 & 0.67 \\ 
181 & 0.72 & 0.72 & 0.79 & 0.65 \\ 
182 & 0.71 & 0.71 & 0.78 & 0.63 \\ \hline \hline 
\end{tabular} 
\end{center} 
\vskip 0.08in 
\caption{The central and mean values
of $\sigma_{W^*}$ (in pb),
along with
the upper and lower bounds derived by varying the scale
and by considering the PDF sets CTEQ4M and MRRS(R1).
The central value is the value midway between the upper and lower
bounds, while the mean is the averaged result of $\sigma_{W^*}$
calculated using the CTEQ4M and MRRS(R1) PDF with the
canonical scale choice discussed in the text.}
\label{schantab}
\end{table}

The $W$-gluon fusion mode has also been studied at NLO
in QCD (order $\alpha_{EW}^2 \alpha_S^1$) \cite{tchan3}.
This production mode has the advantage of a larger cross section
($\sigma_{Wg}$) and a smaller
percentage dependence on $m_t$ than the $W^*$ process,
though the {\it absolute} dependence on
$m_t$ is actually comparable for the two rates
(c.f. Fig \ref{topmassdep}).
This mode is also of interest
because within the SM, it provides a way to directly probe
the partial width of the top quark, $\Gamma(t \ra W^+ b)$,
through the effective-$W$ approximation \cite{effW},
valid at energies much larger than the $W$ mass ($m_W$),
in which the $W$ boson is treated as a parton within the
proton.  Using this approximation, $\sigma_{Wg}$ can
be related to the width $\Gamma(t \ra W^+ b)$ by
the equation \cite{cpy}
\beq
\sigma_{Wg} \simeq \sum_{\lambda = 0, +, -} \;
\int dx_1 \; dx_2 \; f_{\lambda}(x_1) \; b(x_2)
\left[ \frac{16 \pi^2 m_t^2}{\hat s (m_t^2 - M_W^2)} \right]
\Gamma(t \ra W^+_{\lambda} b) \; ,
\label{effWeq}
\eeq
where $x_1 x_2 = \hat s / S$,
$f_\lambda(x_1)$ is the distribution function for
$W$ bosons within the proton carrying momentum fraction
$x_1$ \cite{effW}, $b(x_2)$ is the $b$ quark PDF, and
$\lambda$ is the polarization of the $W$ boson.
Once this partial width has been extracted from a measurement
of $\sigma_{Wg}$, it can be combined with a measurement of the
branching ratio ($BR$) of $t \ra W^+ b$ 
(obtained from examining top decays within
$t \bar t$ production)
to get the top quark's full width ($\Gamma(t \ra X)$, where
$X$ is anything) via the relation \cite{cpy},
\beq
\Gamma( t \ra X) = \frac{\Gamma(t \ra W^+ b)}
{BR( t \ra W^+ b)} \; .
\label{widtheq}
\eeq
This method relies on the fact that within the SM
there are no flavor-changing neutral current (FCNC)
interactions, and the CKM elements $V_{ts}$ and
$V_{td}$ are very small \cite{pdb}; thus the
$t$-channel single top production involves fusion of
only the $b$ parton with a $W^+$ boson\footnote{Assuming
a top mass of $m_t =$ 175 GeV, including the non-zero
contributions from $V_{td} = 0.009$ and $V_{ts} = 0.04$
\cite{pdb} increases the $W$-gluon fusion cross section
by less than $0.5\%$.}

\begin{figure}
\centerline{\hbox{
\psfig{figure=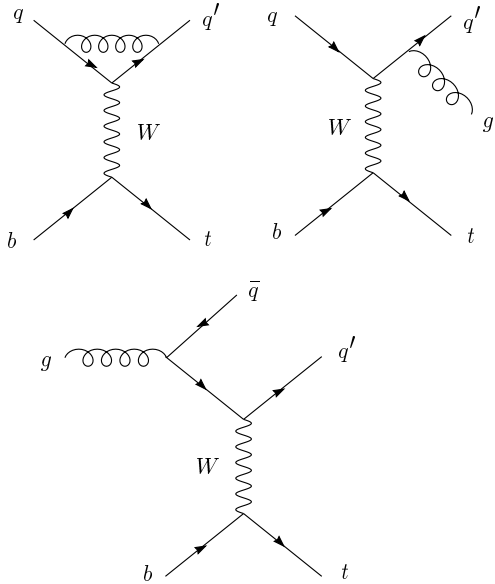,height=3.6in}}}
\caption{ Representative Feynman diagrams for $\alpha_S$
NLO corrections
to the $W$-gluon fusion mode of single top production
coming from the light quark line.}
\label{NLOtl}
\end{figure}

\begin{figure}
\centerline{\hbox{
\psfig{figure=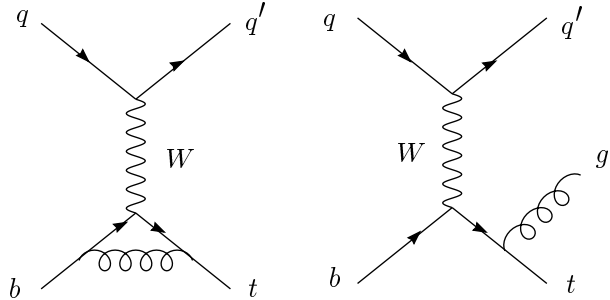,height=2.2in}}}
\caption{ Representative Feynman diagrams for $\alpha_S$
NLO corrections
to the $W$-gluon fusion mode of single top production
coming from the heavy quark line.}
\label{NLOth}
\end{figure}

The major drawback of the $W$-gluon fusion mode
is that it suffers
from a larger theoretical uncertainty due to the uncertainty in the
$b$ quark parton density (through its dependence on
the gluon density).  It has been known for some time
\cite{cpy, dougthesis} that the treatment of the $b$ quark
as a parton must be done carefully, and that a realistic
calculation of the $W$-gluon fusion rate must include
the initial $b$-quark diagram as well as the gluon splitting
diagram (c.f. Figure \ref{tchannelfig})
with the over-lap between the two diagrams when the $b$ quark
becomes collinear with the initial gluon properly
subtracted out, to avoid double counting this
region (which has already been resummed into the
LO rate through the definition of the $b$ PDF).
We refer to the heavy line gluon splitting contribution
as the gluon splitting diagram, minus the kinematic
region where the $b$ parton is collinear to the incoming
gluon.
It
was demonstrated in \cite{cpy,tchan3}
that the corrections due to the
gluon splitting diagram are actually of order
1/${\rm ln}(m_t^2/m_b^2)$ rather than
order $\alpha_S$.  In that
work the genuine order $\alpha_S$ corrections to the light
quark line (c.f. Figure \ref{NLOtl}) and heavy quark line
(c.f. Figure \ref{NLOth}) were also computed.

In order to estimate the theoretical uncertainty in $\sigma_{Wg}$,
we examine its dependence on the choice of scale and PDF.
It is expected that the uncertainties in the gluon
PDF will reflect themselves in a strong dependence of individual
contributions to the NLO cross section from the LO and
gluon splitting pieces (shown in Figure \ref{tchannelfig}),
the $\alpha_S$ corrections to the light quark line
(shown in Figure \ref{NLOtl}), and the $\alpha_S$ corrections
to the heavy quark line (shown in Figure \ref{NLOth}).
However, because of the QCD sum rules governing the PDF,
we expect that the sum of all of these contributions to the
NLO rate will actually show much less sensitivity than
the individual pieces do \cite{cpy, dougthesis}.
In order to illustrate
this point, we examine the individual contributions to the
NLO rate mentioned above for the CTEQ4M, CTEQ4A4, CTEQ4A5
and CTEQ4HJ
PDF\footnote{ We do not use the MRRS(R1) PDF for this analysis,
as it differs from the CTEQ4 PDF in its treatment of the bottom
quark evolution.  Since we are trying to examine the effects of
similar PDF fits for the gluon distribution using different
gluon density parametrizations, we feel it is best to consider
the same $b$ quark evolution scheme in order to avoid confusion
by the different modelling of the $b$ quark mass threshold effects.},
as examples of PDF sets with different shapes for the gluon
density, and in the case of CTEQ4HJ, different
parameterization for the gluon 
density.  The results are presented in
Table \ref{tchanpieces}.  From this comparison, we see
that while individual pieces of the NLO cross section
can vary by about $4\%$,
the total NLO rate varies by less than $1\%$.
Also, further data from Deeply Inelastic Scattering
(DIS) experiments is expected to provide stronger constraints
on the gluon PDF, and thus with time it is expected that we
will better understand and reduce this uncertainty in the
$W$-gluon fusion rate.

\begin{table}[htbp] 
\begin{center} 
\vskip -0.06in
\begin{tabular}{|l||c|c|c|c|}\hline \hline 
Quantity & CTEQ4M (pb) &
CTEQHJ (pb) & CTEQ4A4 (pb) &
CTEQ4A5 (pb) \\ \hline
LO rate & 2.65 & 2.66 & 2.63 & 2.60 \\
1/${\rm ln}(m_t^2/m_b^2)$ & -0.55 & -0.56 & -0.56 & -0.54 \\
Heavy line $\alpha_S$ & 0.30 & 0.30 & 0.31 & 0.31 \\
Light line $\alpha_S$ & 0.04 & 0.04 & 0.04 & 0.04 \\
Total NLO rate & 2.44 & 2.44 & 2.41 & 2.42 \\ \hline \hline
\end{tabular}
\end{center} 
\vskip 0.08in 
\caption{The contributions to $\sigma_{Wg}$ from the LO
rate, the gluon splitting into $b$ diagram, and genuine
$\alpha_S$ corrections to the light and heavy quark lines,
for the CTEQ4M, CTEQ4A4, CTEQ4A5,
and CTEQ4HJ PDF.  Despite the fact
that separate contributions can vary at the $4\%$ level,
the over-all cross section remains stable to
less than $1\%$.}
\label{tchanpieces}
\end{table}

In \cite{tchan3} it was found that the NLO coefficient functions
for the $W$-gluon fusion process contain logarithms which indicate
that the appropriate choice of scale is $\mu = \sqrt{Q^2}$ for the
light line corrections,
and $\mu = \sqrt{m_t^2 + Q^2}$ for the heavy line corrections.
In that work it was argued that due to the strong analogy between
$W$-gluon fusion and the DIS process by which most of the
experimental understanding of the PDF is derived, it is not
necessary to vary the scale of the light line PDF to estimate
the uncertainty in the calculation.
However, in order to be conservative, we examine the result of
simultaneously
varying $\mu$ between $2 \sqrt{Q^2}$ and $\sqrt{Q^2} / 2$ in
the light line quantities, and between $2 \sqrt{m_t^2 + Q^2}$
and $\sqrt{m_t^2 + Q^2} / 2$ in the heavy line quantities,
for the CTEQ4M and MRRS(R1) PDF\footnote{We have also examined
the predictions of the CTEQ4A4, CTEQ4A5 and CTEQ4HJ
PDF for $\sigma_{Wg}$ in the range of top masses
we are considering.  We find that all of these PDF give
results for $\sigma_{Wg}$ that differ from CTEQ4M by
less than $1\%$ (Table \ref{tchanpieces} illustrates
this for $m_t = 175$ GeV).  Thus, the limits on the
range of $\sigma_{Wg}$ we derive are consistent with
the full set of PDF MRRS(R1), CTEQ4M, CTEQ4HJ, CTEQ4A4,
and CTEQ4A5.}.
These PDF differ in that
the MRRS(R1) set has a different treatment of the charm and
bottom PDF evolution near their mass thresholds.  This is not
expected to be important in the heavy
line contribution to $W$-gluon fusion,
where $\mu = \sqrt{m_t^2 + Q^2} \gg m_b , m_c$.
However, it is also necessary to modify the usual
$\overline{\rm MS}$
subtraction term for the gluon splitting diagrams involving
the $c$ and $b$ quarks
in order to
correctly deal with the way the $b$ PDF is derived in the MRRS
formalism \cite{mrsa}.
{}From this we arrive at the
theoretical estimate for the total $W$-gluon fusion rate, as
well as the upper and lower bands for its variation,
and the central value between these bounds, in the same
way that those for the $W^*$ cross section were obtained.
This allows us to examine the correlated effect to the
$W$-gluon fusion rate due to PDF choice, scale, and $m_t$.
Once again, we see that the central and mean
values are very close, indicating that the scale
choices are appropriate.
These
results, for various values of the top mass are presented
in Table \ref{tchantab} (and shown graphically in Figure
\ref{topmassdep}).
In obtaining these results, we
have included the full effects of the CKM matrix in the light
quark vertex, but have ignored them in the heavy quark vertex.
This prescription effectively amounts to assuming $V_{tb} = 1$,
since the contributions from $V_{td}$ and $V_{ts}$ are negligible
anyway (see footnote 4).
Using the variable $\Delta m_t$ defined above, one may parameterize
these curves (in units of pb) as,
\begin{eqnarray} 
\sigma^{\rm mean}_{Wg}(\Delta m_t) = 2.35 - 
(3.2 \times 10^{-2}) {\Delta m_t} -  
(2.4 \times 10^{-3}) {\Delta m_t}^2 - \nonumber \\
(3.2 \times 10^{-4}) {\Delta m_t}^3 +
(8.0 \times 10^{-5}) {\Delta m_t}^4 \nonumber \\
\nonumber \\
\sigma^{\rm central}_{Wg}(\Delta m_t) = 2.37 - 
(3.1 \times 10^{-2}) {\Delta m_t} -
(1.5 \times 10^{-3}) {\Delta m_t}^2 - \\
(3.0 \times 10^{-4}) {\Delta m_t}^3 +
(5.4 \times 10^{-5}) {\Delta m_t}^4 \nonumber \\
\nonumber \\
\sigma^{\rm upper}_{Wg}(\Delta m_t) = 2.55 - 
(2.3 \times 10^{-2}) {\Delta m_t} -  
(4.5 \times 10^{-3}) {\Delta m_t}^2 - \nonumber \\
(8.3 \times 10^{-4}) {\Delta m_t}^3 +
(1.6 \times 10^{-4}) {\Delta m_t}^4 \nonumber \\
\nonumber \\
\sigma^{\rm lower}_{Wg}(\Delta m_t) = 2.19 - 
(3.1 \times 10^{-2}) {\Delta m_t} -  
(1.6 \times 10^{-3}) {\Delta m_t}^2 - \nonumber \\
(1.9 \times 10^{-4}) {\Delta m_t}^3 +
(4.8 \times 10^{-4}) {\Delta m_t}^4 \nonumber \;.
\end{eqnarray}

\begin{table}[htbp] 
\begin{center} 
\vskip -0.06in
\begin{tabular}{|l||c|c|c|c|}\hline \hline 
$m_{t}$ (GeV) & $\sigma_{Wg}$ (mean) (pb) &
$\sigma_{Wg}$ (central) (pb) &
$\sigma_{Wg}$ (upper) (pb) & $\sigma_{Wg}$ (lower) (pb) \\ \hline
170 & 2.54 & 2.56 & 2.76 & 2.36 \\
171 & 2.48 & 2.50 & 2.67 & 2.31 \\     
172 & 2.44 & 2.46 & 2.62 & 2.28 \\ 
173 & 2.41 & 2.43 & 2.59 & 2.25 \\ 
174 & 2.38 & 2.40 & 2.57 & 2.22 \\ 
175 & 2.35 & 2.37 & 2.55 & 2.19 \\
176 & 2.32 & 2.34 & 2.52 & 2.16 \\ 
177 & 2.28 & 2.30 & 2.48 & 2.12 \\ 
178 & 2.23 & 2.26 & 2.43 & 2.08 \\ 
179 & 2.18 & 2.22 & 2.37 & 2.04 \\ 
180 & 2.14 & 2.18 & 2.32 & 2.00 \\ 
181 & 2.11 & 2.14 & 2.28 & 1.96 \\ 
182 & 2.09 & 2.11 & 2.27 & 1.94 \\ \hline \hline 
\end{tabular} 
\end{center} 
\vskip 0.08in 
\caption{The central value of $\sigma_{Wg}$, along with the 
mean value (for the canonical scale choice described in the text)
and the upper
and lower bounds derived by varying the scale and by considering the PDF
sets CTEQ4M and MRRS(R1).} 
\label{tchantab}
\end{table}

Studying single top production at the Tevatron Run II
(and beyond)
is expected to yield the first direct measurements of $V_{tb}$,
the top - bottom CKM matrix parameter \cite{cpy}.
While the assumption of three
generations, combined with unitarity considerations, allows one
to constrain $V_{tb}$ to within a few percent
of $V_{tb}$ = 0.9991 \cite{pdb}, this bound
disappears if one allows for more than three generations.
Thus, it
is important to directly measure $V_{tb}$,
as it may provide a clue
concerning the existence of a
fourth generation of fermions which mixes with the
third family through the CKM matrix.  A method to
accomplish this
at the Tevatron Run II has been proposed, using the
single top $W$-gluon fusion \cite{cpy} and $W^*$ \cite{wvtb}
rates.  The method
relies upon $t \bar t$ production to measure the branching ratio
for $t \ra b W$, and then uses this information in conjunction with a
measurement of single top production (through either mode),
which is proportional to
$|V_{tb}|^2$, to extract $|V_{tb}|$ independently from the top quark
decay mode.
However, as we shall show in Section \ref{resonance}, the presence
of new physics in the top quark sector can result in a large
modification of either $\sigma_{W^*}$
or $\sigma_{Wg}$,
which could distort the value
of $V_{tb}$ obtained in this way.  We will return to this question
in Section \ref{vtb}, after examining how an additional heavy
resonance or modified top quark interactions can affect
$\sigma_{Wg}$ and $\sigma_{W^*}$.

In attempting to estimate the total uncertainty in the cross
sections of $\sigma_{W^*}$ and $\sigma_{Wg}$ due to both
theoretical and statistical uncertainties, we assume
a top mass of $m_t = 175$ GeV, and that the actual
physical values of the cross sections are the same
as the mean values at $m_t =$ 175 GeV.  We have seen
from the results in Tables \ref{schantab} and
\ref{tchantab} that the effect on the uncertainty of
the cross sections from the
uncertainties due to the scale
dependence, PDF choice, and uncertainty in the top
mass\footnote{We assume that with 2 \fb of integrated
luminosity or more, the uncertainty in the top mass
will be $\delta m_t = \pm 2$ GeV \cite{tev2000}.}
are actually correlated with one another.
For the theoretical uncertainty, we see from Tables
\ref{schantab} and \ref{tchantab} that for a top mass
of $m_t = 175 \pm 2$ GeV, the expected values and
uncertainties of $\sigma_{W^*}$ and $\sigma_{Wg}$
are,
\begin{eqnarray}
\sigma_{W^*} = 0.84 \; {\rm pb} \pm 15\% \\
\sigma_{Wg}  = 2.35 \; {\rm pb} \pm 10\% \nonumber \; .
\end{eqnarray}

However, this theoretical uncertainty is not correlated with the
statistical uncertainty in the measurement, and thus
these two fractional
uncertainties may be added in quadrature to
arrive at the total theoretical and statistical uncertainty
in $\sigma_{W^*}$ and $\sigma_{Wg}$.
In deriving the projected statistical uncertainty, we
assume a top quark decay of \tsmdecay with
$\ell = e$ or $\mu$, and thus include a
branching ratio of $2 / 9$ for this decay mode.
In order to
include an estimation of
the detection efficiencies at the Tevatron Run II, we include
the LO efficiency factors of
$9\%$ for the $W^*$ production mode\footnote{This efficiency for
detecting the $W^*$ mode includes a factor of $36\%$
for double $b$-tagging when both $b$ quarks have
$p_{Tb} \geq 20$ GeV and $|\eta_b| \leq 2$, 
based on an estimated $60\%$
efficiency for single $b$-tagging of a $b$ quark in
this kinematic region.}
\cite{schan1} and
$33\%$ for the $W$-gluon fusion mode\footnote{Thise estimate
for the $W$-gluon fusion detection efficiency
include a $60\%$ single $b$-tagging efficiency
when the $b$ quark has
$p_{Tb} \geq 35$ GeV and $|\eta_b| \leq 2$.} \cite{dougthesis}.
This is some-what crude, as the NLO results include kinematic
configurations where an additional jet is present in the
final state\footnote{The theoretical uncertainty in the
production rate may also depend on the kinematic cuts
imposed to enhance the signal-to-background ratio.
This requires a study on the detailed distributions of
the final state particles in the single top events,
and is thus beyond the scope of this work.},
and thus the actual efficiencies are probably slightly
different from these estimates.  Nonetheless, when considering
new physics effects in single top production,
we will not rely
on detailed studies of the final state kinematics in our
analysis, and thus this should not have a large effect on our
conclusions.
From these results we derive the projected statistical uncertainty
in the measured
cross sections for an integrated luminosity of 2, 10, and 30
\fb.  The results, including the theoretical uncertainties
derived above, and the total over-all uncertainty obtained
by adding the fractional uncertainties in quadrature, are
summarized in Table \ref{uncertaintytab}.
It is also interesting to examine the theoretical
uncertainties derived by varying the top mass,
scale, and PDF, without taking into account
the correlations in the dependence of the cross sections
on these quantities, though as we have argued above, this
is not the correct way to estimate the full theoretical
uncertainty.  Assuming the canonical scale choices for
each mode described above, $m_t = 175$ GeV, and the
CTEQ4M PDF, we vary each of these quantities individually
by the same range used in our correlated analysis above,
and examine the effect on the cross section.  The resulting
fractional uncertainty in the cross sections is presented in
Table \ref{quaduntab}.  From this we see clearly that if
one considers the separate contributions to the theoretical
uncertainty as uncorrelated (and thus adds them in quadrature),
one under-estimates the true uncertainty shown in Table
\ref{uncertaintytab}.  We also see that the correlations
in these quantities are some-what stronger for
$\sigma_{W^*}$ than for $\sigma_{Wg}$.
We note that
our estimate of the theoretical
uncertainty in the $W^*$ rate is some-what
more pessimistic than the estimate of
$\pm 10\%$ in \cite{schan2}, where the
complete
correlation in the theoretical uncertainties was not taken
into account.

\begin{table}[htbp] 
\begin{center} 
\vskip -0.06in
\begin{tabular}{|l||c|c|}\hline \hline 
Quantity & $\delta \sigma_{W^*}$ &
$\delta \sigma_{Wg}$ \\ \hline
Theoretical & $\pm 15\%$ & $\pm 10\%$ \\
Statistical (2 \fb) & $\pm 17\%$ (34 Events) & $\pm 5\%$ (345 Events) \\
Statistical (10 \fb) & $\pm 8\%$ (168 Events) & $\pm 2\%$ (1723 Events) \\
Statistical (30 \fb) & $\pm 4\%$ (504 Events) & $\pm 1\%$ (5170 Events) \\
\hline
Total (2 \fb) & $\pm 23\%$ & $\pm 11\%$ \\
Total (10 \fb) & $\pm 17\%$ & $\pm 10\%$ \\
Total (30 \fb) & $\pm 16\%$ & $\pm 10\%$ \\
\hline \hline 
\end{tabular} 
\end{center} 
\vskip 0.08in 
\caption{The correlated theoretical uncertainty from scale, PDF,
and uncertainty in $m_t$ for $\sigma_{W^*}$ and $\sigma_{Wg}$,
as well as the projected statistical uncertainty for various
integrated luminosities assuming a
semi-leptonic top decay into
an electron or muon, and the detection efficiencies of
$9\%$ for the $W^*$ process and $33\%$ for the $W$-gluon
fusion process obtained from LO studies.  A top mass of
175 GeV with an uncertainty of $\pm 2$ GeV is assumed.}
\label{uncertaintytab}
\end{table}

\begin{table}[htbp] 
\begin{center} 
\vskip -0.06in
\begin{tabular}{|l||c|c|}\hline \hline 
Quantity Varied& $\delta \sigma_{W^*}$ & $\delta \sigma_{Wg}$ \\ 
\hline
$\delta m_t$  & $\pm 6\%$ & $\pm 3\%$ \\
Scale                    & $\pm 5\%$ & $\pm 4\%$ \\
PDF                      & $\pm 6\%$ & $\pm 7\%$ \\
\hline
Total added in Quadrature  & $\pm 10\%$ & $\pm 9\%$ \\
Total added linearly       & $\pm 18\%$ & $\pm 14\%$ \\
Correctly Correlated Total & $\pm 15\%$ & $\pm 10\%$ \\
\hline \hline
\end{tabular} 
\end{center} 
\vskip 0.08in 
\caption{The uncorrelated theoretical uncertainty from scale, PDF,
and uncertainty in $m_t$, for $\sigma_{W^*}$ and $\sigma_{Wg}$,
and the totals obtained from adding these uncertainties in
quadrature (assuming they are correlated), linearly (assuming
$100\%$ correlation), and the result including the true
correlation from Table 4.}
\label{quaduntab}
\end{table}

From these results, we learn that
for 2 \fb of integrated luminosity, the $W$-gluon fusion
sub-process has an uncertainty which is 
about a factor of 2 better than
that of the
$W^*$ mode.  This is because the $W^*$ mode's larger
sensitivity to the top mass, and the strong correlations
between this uncertainty and those arising from the PDF
and scales,
and smaller cross section
(and detection efficiency)
act to compensate for the larger uncertainty in the $W$-gluon
fusion mode due to the parton distribution functions.  For the
larger data samples considered, the improvements in available
statistics allow the
uncertainties of the two rates to become comparable, although
the $\sigma_{W^*}$ uncertainty remains dominated by that mode's
larger percentage dependence on the top mass, and thus is
larger.
These results must be considered with some care, as no
systematic experimental uncertainties are included, and these
are likely to be different for the two production modes.

\section{New Physics in the Form of an Additional Heavy Resonance}
\label{resonance}
\indent\indent

\begin{figure}
\centerline{\hbox{
\psfig{figure=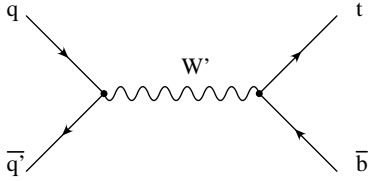,height=1.0 in}}}
\caption{Feynman diagram showing how
an additional heavy charged vector particle ($W'$)
can contribute to the $s$-channel process of
single top production.}
\label{wpdiags}
\end{figure}

\begin{figure}
\centerline{\hbox{
\psfig{figure=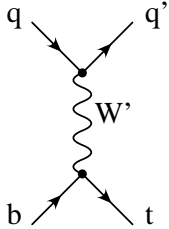,height=1.3 in}}}
\caption{Representative Feynman diagram showing how
an additional heavy charged vector particle ($W'$)
can contribute to the $W$-gluon fusion mode of
single top production.}
\label{wpdiagt}         
\end{figure}

One possible form of new physics in the top quark sector is an
additional resonance, beyond those required by the SM, which
couples to the top quark.  In particular, a resonance with
electric charge $Q = +1$ could couple
to the top and bottom quarks, and
thus could contribute to single top production. 
Generically, a heavy vector boson with charge $Q$ = +1
(which we shall refer to as a
$W'$)
can affect the rate of single top production 
(in either the $s$ or $t$ channel processes) by
contributing additional diagrams in which the $W'$ is
exchanged, such as those shown in Figures \ref{wpdiags}
and \ref{wpdiagt}.  Because the
initial and final states are the same for both the
$W$ exchange and $W'$ exchange diagrams, they can interfere
at the amplitude level, and thus the effect of the $W'$
could either raise or lower the single top cross sections,
depending on the relative sign of the couplings between the
$W$ and the fermions, and the $W'$ and the fermions.

We expect that the two modes of single top production
will show a very different sensitivity to the presence of
a $W'$.  The $s$-channel $W^*$ process can show a large
sensitivity, because the time-like momentum of the
exchanged $W'$ can be close to on-shell, thus providing
an enhancement from the $W'$ propagator in the $s$-channel
matrix element (c.f. Figure \ref{wpdiags}).  This also
suggests that one way to look for the presence of a
$W'$ is to examine the distribution of $\sigma_{W^*}$
with respect to the invariant mass of the $t$ $\bar b$
system ($M_{tb}$)\footnote{There is an ambiguity in reconstructing
$M_{tb}$ when the top decays semi-leptonically, as we
are assuming in this work, because the final state neutrino
is not observed, and thus the component of its momentum along
the beam axis is not measured.  However, it is possibly to
reconstruct the top's momentum in a statistical way
\cite{schan2, tchan1, dougthesis}, thus allowing a similar
reconstruction of $M_{tb}$.},
for bumps from the $W'$ resonance, or for
signs of a $W'$ tail.
In fact the authors of \cite{schan1} found that in
order to distinguish the $W^*$ signal from the large
$W b \bar b$ background at the Tevatron, it is necessary to
require a large invariant mass between the $b$ and $\bar b$
quarks in the final state, $M_{b\bar b} \geq$ 110 GeV.  Since
the final state $b$ quark
results from the decay of the top, this is
equivalent to requiring that the invariant mass
$M_{tb}$ is large, and thus the 
observed $W^*$ signal comes mostly from the region
of $M_{tb}$ which is most sensitive to the effect of the $W'$.
However, if the mass
of the $W'$ is too large, or its width too broad, the effect in
$M_{tb}$
can be washed out at the Tevatron, preventing one
from identifying the presence of the $W'$ in this way.

On the other hand, the $t$ channel process
requires a space-like momentum for the $W'$ boson
(c.f. Figure \ref{wpdiagt}), and thus
can never experience this type of resonant propagator
enhancement.  The additional
$t$-channel diagram containing the $W'$ (c.f.
Figure \ref{wpdiagt})
will be suppressed by $1 / M^2_{W'}$.
Furthermore, the kinematic distribution of the invariant mass
$M_{tq'}$, where $q'$ is the light quark in the final state
(c.f. Figure \ref{tchannelfig}), shows
the same characteristics as the SM distribution, and thus does
not provide much additional information beyond that contained
in the total rate, $\sigma_{Wg}$.
The $t W^-$ process is also 
not expected to show an effect
from the presence of a $W'$, because the large mass of the $W'$
forbids direct production of a top and a $W'$.
Similarly, the large mass of the $W'$ will
prevent it from affecting top quark decays since $M_{W'}$ must
be greater than $m_{t}$ in order to respect current limits on
a $W'$ mass \cite{pdb}.

For the purposes of illustrating
these points, we will consider the (light)
top-flavor model \cite{tf, tf1, tf2},
in which the third family of fermions under-goes a separate SU(2)
weak interaction from the first two families\footnote{A
comprehensive introduction to this model,
may be found in \cite{tf1}.  We are using the same conventions
as were used in that work.}.
In the top-flavor model the
over-all gauge symmetry is
SU(2$)_{h} \times$ SU(2$)_{l} \times$ U(1$)_{Y}$, and thus
there are three additional weak bosons ($W'^{\pm}$ and $Z'$).
The first and second generation fermions transform under
SU(2$)_{l}$, while the third generation fermions transform
under SU(2$)_{h}$.  A set of scalar fields transforming under
both SU(2$)_{l}$ and SU(2$)_{h}$ acquire a vacuum expectation
value (v.e.v.), $u$,
and break the symmetry to SU(2$)_{l + h} \times$
U(1$)_{Y}$.  From here the usual electro-weak symmetry breaking
can be accomplished by introducing a scalar doublet which acquires
a v.e.v. $v$, further breaking the gauge symmetry to U(1$)_{EM}$.
The model is parameterized by two quantities, $x = u / v$, and
$\sin^{2} \phi$, which characterizes the mixing between the heavy
and light SU(2) gauge couplings.  At leading order, the heavy bosons
are degenerate in mass,
\beq
{M^2}_{Z', W'} = {M_0}^2 \left( \frac{x}{\sin^2 \phi \cos^2 \phi}
+ \frac{\sin^2 \phi}{\cos^2 \phi} \right) ,
\eeq
where ${M_0}^2 = \frac{ e^2 v^2 }{4 \sin^2 \theta \cos^2 \theta}$,
and
$e$ and $\theta$ are the U(1$)_{EM}$ coupling and the weak
mixing angle, respectively.
We can thus parameterize the model by the
heavy boson mass, $M_{Z'}$, and the
mixing parameter\footnote{For $\sin^2 \phi \leq$ 0.04, the
third family fermion coupling to the heavy gauge bosons can
become non-perturbative.  Thus we restrict ourselves to considering
$0.95 \geq \sin^2 \phi \geq$ 0.05. (c.f. Ref. \cite{tf1}).},
$\sin^2 \phi$.
Low energy data requires that the mass
of these heavy bosons, $M_{Z'}$,
be greater than about $1.1$ TeV \cite{tf1}.
The impact of this type of model on the $W^*$
sub-process\footnote{In this section, we use "$W^*$ process"
to denote $q \bar{q'} \ra W, W' \ra t \bar b$.  The results
we present will also include single anti-top production
through $q \bar{q'} \ra W, W' \ra \bar t b$.}
of single top production through the observable
$R_{\sigma} = \frac{\sigma(q \bar{q'} \ra W^*, W' \ra tb)}
{\sigma(q \bar{q'} \ra W, W' \ra \ell \nu)}$
was considered in \cite{es}.  The conclusions drawn there
about the possibility of detecting top-flavor
at the Tevatron Run II by observing single top
production are similar to ours.

\begin{figure}
\centerline{\hbox{
\psfig{figure=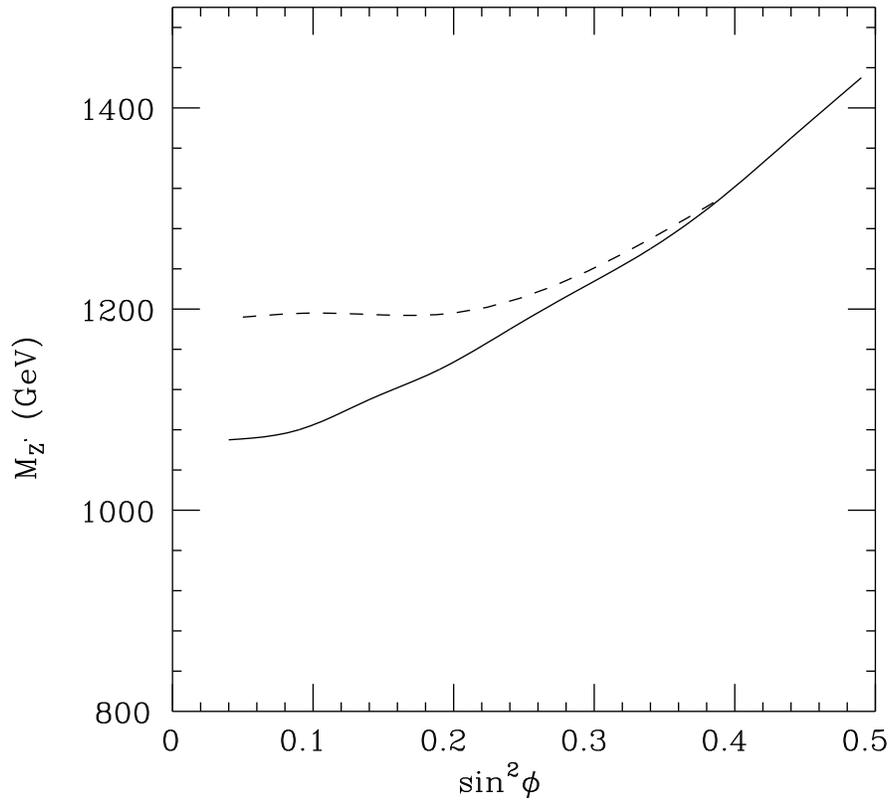,height=4.0in}}}
\caption{The region of top-flavor parameter space allowed
by low energy data (above the solid curve), and modifying the
$\sigma_{W^*}$ rate by at least $15\%$ (below the dashed
line).}
\label{tflim}
\end{figure}

To illustrate these ideas concerning the effect of a
$W'$ on the single top production rates, we examine the
prediction of the top-flavor
model for single-top production 
at NLO in both the $s$ and $t$
channels, in the region of
top-flavor parameter space not ruled out by low energy data.
We assume a top mass of $m_{t} = 175$ GeV.
We find that in this region of parameter space,
the $W^*$ production mode is generally increased by about $15\%$,
for heavy boson masses in the region around
$M_{Z'} = 1.1$ TeV (c.f. Figure \ref{tflim}).
We also examine the effects of this model
to the $W$-gluon fusion rate.  As expected, we find that the
$W$-gluon fusion cross section 
is quite insensitive to the presence of the $W'$,
deviating from the SM value by less than $2\%$ in the
entire range of allowed parameters.

\begin{figure}
\centerline{\hbox{
\psfig{figure=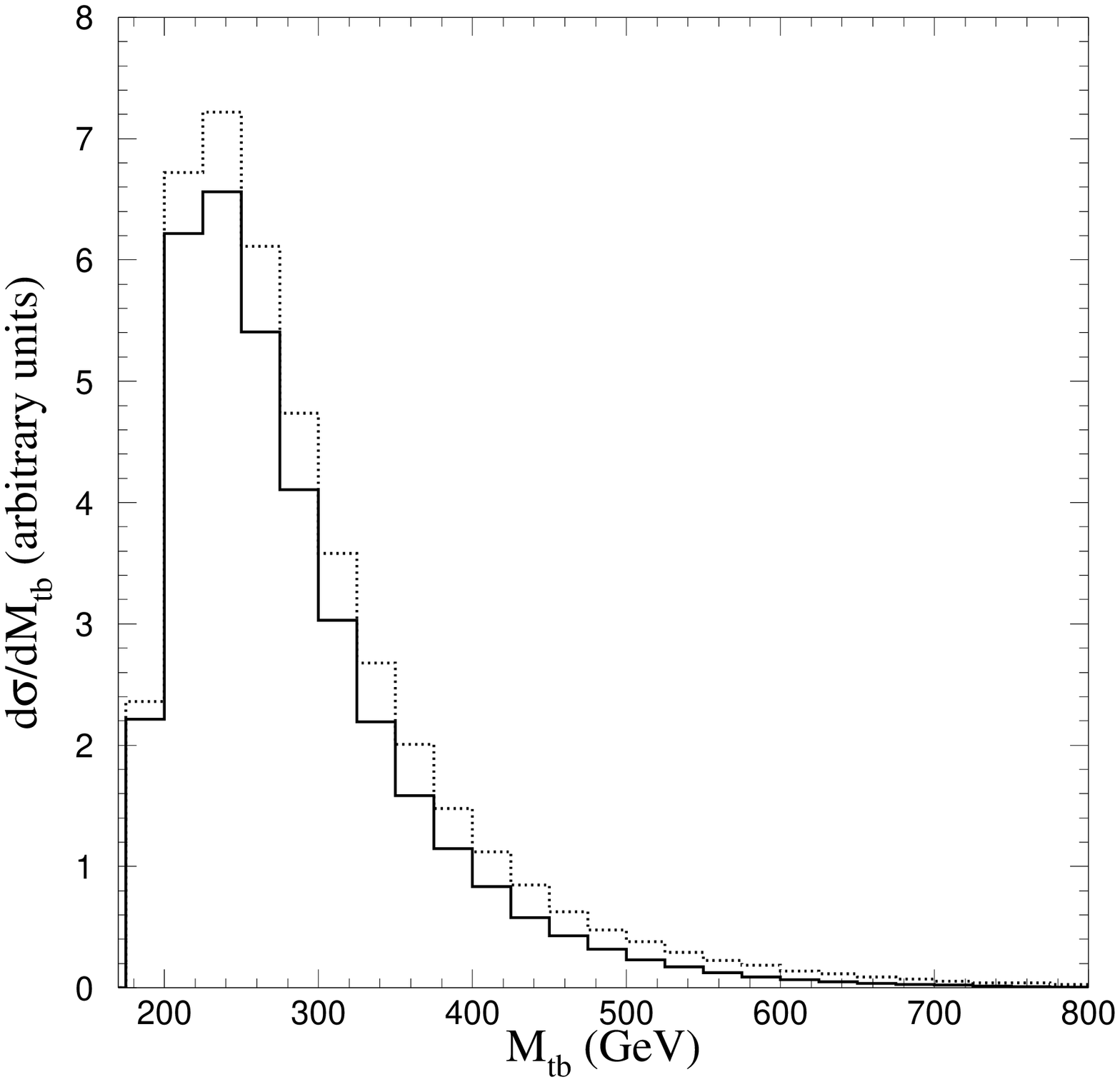,height=4.0in}}}
\caption{The distribution of $\frac{d\sigma_{W^*}}{dM_{tb}}$
for the SM (solid curve), and the top-flavor model (dotted
curve) with $\sin^2 \phi = 0.05$ and $M_{Z'} = 1.2$ TeV.
(The vertical scale of the histogram is in arbitrary units.)
Both calculations assume a top mass of $m_t = 175$ GeV.}
\label{mtb}
\end{figure}

As mentioned previously, one could look for
the effect of the heavy resonance in
the distribution of $M_{tb}$ for $W^*$ single-top
events.  With enough statistics, it
is likely that examining this distribution could allow
one to further constrain the top-flavor model for regions
of parameter space where the $W'$ is not too heavy, and its
width is not too broad.
However, in the region shown in Figure \ref{tflim}
the effect is washed out
by the heavy $W'$ mass, and its broad width.
Further, with limited statistics it may not be practical
to consider distribution of $M_{tb}$.
To illustrate this point, in
Figure \ref{mtb} we show $\frac{d \sigma_{W^*}}{d M_{tb}}$
for the SM as well as for a typical choice
of allowed top-flavor
parameters which modify the total rate $\sigma_{W^*}$ by
at least $15\%$.  From this result one can see that
the $M_{tb}$ effect can be
washed out by the large mass, and broad width of
the $W'$.

These results reveal a limitation in relying solely on the
$W^*$ process to measure $V_{tb}$.  An additional charged
resonance can have a large effect on the $W^*$ cross section,
which could lead one to mis-measure $V_{tb}$.  While for
the specific example of the top-flavor model, the effect
of the $W'$ was to increase the $W^*$ cross section,
and thus would be likely to lead to a measurement of 
$V_{tb} \geq$ 1 (which would in itself signal that there
is a problem in using the $W^*$ rate to measure $V_{tb}$),
in a general model where the $W'$ couplings to the
fermions have an arbitrary sign relative to the $W$ couplings,
the amplitudes for $W$ and $W'$ can interfere, leading to a
decrease in $\sigma_{W^*}$ (and thus decreasing the value of
$V_{tb}$ one would extract from such a measurement).
This could
lead one to conclude, erroneously, that there
was evidence for a fourth generation of fermions mixed
with the third generation via the CKM matrix.  Thus,
to be sure one is measuring $V_{tb}$ accurately, it is
not enough to rely completely on $\sigma_{W^*}$.

Other types of heavy resonance can be added to the SM, and analyzed in
a similar fashion.  If the additional particle has charge +1
(such as the $W'$ considered here), and
couples to both heavy and light quarks, it can contribute to 
the single top rate
through the $W^*$ mode, but should not have a large effect
on the $W$-gluon fusion process because of suppression by
the heavy mass and space-like momentum.  Other types of
new particles, such as e.g., those found in
$R$-parity-conserving
super-symmetry (SUSY), are unlikely to affect
$\sigma_{W^*}$ or $\sigma_{Wg}$ in a large way, though
they could modify the top's decay width and branching ratios.
For example, a charged Higgs, $H^\pm$ could modify the
top width by allowing decays such as $t \ra H^+ b$,
but would not affect the $W^*$ or $W$-gluon fusion
rates because the Higgs couples very weakly to the light
fermions.  This example provides an illustration
of the sense in which the $W$-gluon fusion process is a
measure of the top quark's 
partial width, $\Gamma(t \ra W^+ b)$.  In the case of an
additional $W'$, the top's width was not modified, and we
also saw that the $W$-gluon fusion rate was not modified.
In the case in which there is a charged Higgs $H^+$, the
top's total width is modified because the decay $t \ra H^+ b$
becomes allowed (assuming $m_t > M_{H^+}$),
but the partial width $\Gamma(t \ra W^+ b)$ is unchanged,
and so we find that
the $W$-gluon fusion process is not
modified. Thus it
still can be used to extract the partial
width $\Gamma( t \ra W^+ b)$ via Equation (\ref{effWeq}).
One could
then use Equation (\ref{widtheq}) and $BR(t \ra W^+ b)$,
extracted from the $t \bar t$ event sample,
to calculate the full width of the top quark.

\section{New Physics in the Form of Modified Top Quark Interactions}
\label{width}
\indent\indent

It is also possible that the top quark may couple differently to
light particles from what is predicted by the SM
\cite{cmy, Ztc, tcgam, tcg, Zbb}.  As an example to illustrate
the general features of this type of new physics, we will consider
a flavor-changing neutral current (FCNC) coupling the top to the
charm quark and the $Z$ boson.  To introduce such
an interaction, we
consider
the SM Lagrangian as an effective theory,
allow the SU(2$)_L \times$ U(1$)_Y$ gauge symmetry
to be realized nonlinearly \cite{nonlinear}, and include
the dimension four FCNC term \cite{Ztc},
\beq
\Delta {\cal L}^{eff} = \frac{e \; \kappa^Z_{tc}}
{2 \sin {\theta}_W \cos {\theta}_W}
Z^{\mu}  \; \left[ \bar{t} \; \gamma_{\mu} \; c
+ \bar{c} \; \gamma_{\mu} \; t \right] \;,
\eeq
where $\kappa^Z_{tc}$ parameterizes the strength of the anomalous
term for $Z$-$t$-$c$ coupling, and is
assumed to be real.  As shown in \cite{Ztc},
low energy data requires $|\kappa^Z_{tc}| \leq$
0.29 (at $95\%$ C.L.).

\begin{figure}
\centerline{\hbox{
\psfig{figure=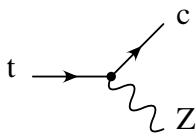,height=0.8in}}}
\caption{Feynman diagram indicating how an anomalous
$Z$-$t$-$c$ vertex can produce the decay $t \ra Z c$.}
\label{tczdecay}
\end{figure}

\begin{figure}
\centerline{\hbox{
\psfig{figure=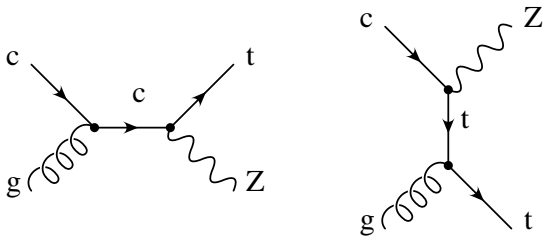,height=1.3in}}}
\caption{Feynman diagrams indicating how an anomalous
$Z$-$t$-$c$ vertex can produce the additional mode
of single top production, $g c \ra t Z$.}
\label{tzprod}
\end{figure}

This form of new physics will
allow new decay modes for the top quark, and thus will modify
the top quark's full decay width, $\Gamma(t \ra X)$,
and cause its branching ratio $BR(t \ra b W^+)$ to deviate
from the SM prediction of $\sim 100\%$, by allowing new decay
modes such as, e.g. $t \ra Z c$, through diagrams such as that
pictured in Figure \ref{tczdecay}.
Also, a FCNC
cannot modify the rate of $t W^-$ production, though
it does allow a new mode of single top production by allowing
a $t Z$ final state (c.f. Figure \ref{tzprod}).
However, the measured rate of $t W^-$ is not sensitive to
a $Z$-$t$-$c$ vertex, and thus
does not provide information about the
strength of such an anomalous coupling.

\begin{figure}
\centerline{\hbox{
\psfig{figure=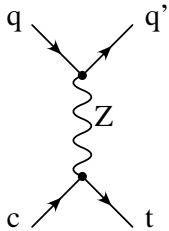,height=1.3 in}}}
\caption{Representative Feynman diagram showing how
an anomalous $Z$-$t$-$c$ coupling can contribute
to the $t$-channel $W$-gluon fusion process of
single top production.}
\label{tchanztc}
\end{figure}

Should the top's
couplings involving the partons of the proton
(i.e. the $u$, $d$, $s$, $c$, and $b$ quarks, and the gluons)
be modified, this may also have an impact on the production of
single tops in the $t$-channel mode
through diagrams such as that shown in Figure
\ref{tchanztc}, where an incoming charm parton experiences
a flavor-changing neutral current involving a $Z$ boson,
producing a top quark and a forward jet in the final state.
Since the distribution of charm quarks in the proton is
larger than distribution of bottom quarks relevant for the
SM $t$-channel production mechanism (c.f. Figure
\ref{tchannelfig}), this contribution may be visible even if
the FCNC $Z$-$t$-$c$ vertex coupling strength is
not large.
In this case we also see that in a sense
the $W$-gluon fusion process is sensitive to the top quark width;
a modification of the top's couplings to the light particles
found as partons inside the proton
will result in a change of the top decay
branching ratios, and
the rate of $W$-gluon fusion production.  Both
properties of the top are thus
sensitive to the same type of new physics.
However, in the case in which a FCNC is present,
it is no longer possible to
use the $W$-gluon fusion process to directly measure the
partial width $\Gamma(t \ra W^+ b)$, as could be done for
a purely SM top, or in a scenario in which an additional
heavy charged resonance is present.

At the
Tevatron Run II, we estimate the effect on the $W$-gluon
fusion cross section to be $\delta \sigma_{Wg}$ = 1.36 pb for
$\kappa^Z_{tc} = 0.29$, and $m_t$ = 175 GeV.
This would constitute a $3 \sigma$
deviation from the expected SM cross section, and thus it is
possible to use single top production to constrain the size
of such a FCNC term at the $99\%$ C.L.
Requiring that no $3 \sigma$ deviation
in the $W$-gluon fusion rate is observed allows one to derive
the constraint $|\kappa^Z_{tc}| \leq$ 0.22 for $L = 2$ \fb,
or $|\kappa^Z_{tc}| \leq$ 0.21 with 10 \fb of
integrated luminosity.  In \cite{Ztc}, it was found
that by studying the $BR$ of the
decay $t \ra Z c$, it is possible to
constrain (at the $99\%$ C.L.) $|\kappa^Z_{tc}| \leq$ 0.3
with $L =$ 2 \fb, or $|\kappa^Z_{tc}| \leq$ 0.16 with
$L = $ 10 \fb.  Thus,
the $W$-gluon fusion rate provides a
comparable means to explore $\kappa^Z_{tc}$ to
that provided by studying $t \ra Z c$.  However, in order to
use the branching ratio $BR(t \ra Z c)$ to constrain
$\kappa^Z_{tc}$ as was done in \cite{Ztc},
it is necessary to 
make the strong assumption that there is no
other new physics
modification to the top decays besides $t \ra Z c$,
so that one can directly convert the branching ratio
$BR(t \ra Z c)$ into the partial width $\Gamma(t \ra Z c)$.
Since the anomalous contribution to the $W$-gluon fusion
cross section is proportional to $|\kappa^Z_{tc}|^2$,
single top production provides
a way to constrain $\kappa^Z_{tc}$ without relying on
strong assumptions about the possibility of other forms
of new physics in the top quark sector.

\begin{figure}
\centerline{\hbox{
\psfig{figure=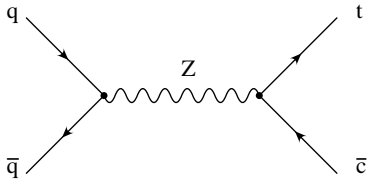,height=1.0 in}}}
\caption{Representative Feynman diagram showing how
an anomalous $Z$-$t$-$c$ vertex can contribute
to the $s$-channel $W^*$ production mode of
single top production.}
\label{schanztc}
\end{figure}

It is also possible for this FCNC vertex to modify the rate
of $W^*$ single top production through diagrams such as
that shown in Figure \ref{schanztc}.  However, this contribution
is unlikely to significantly affect the $W^*$ rate measured at
the Tevatron, because in \cite{schan1} it was found that in
order to isolate the $W^*$ mode from the large $Wjj$ background
and from the $W$-gluon fusion mode, it is necessary to tag
both the $ \bar b$ produced when the $W^*$
decays and the $b$ produced from the top decay,
$t \ra W b$. Thus if another type of quark is produced with the top
(such as the charm quark pictured in Figure \ref{schanztc}),
the additional events will not be included in the $W^*$
measurement\footnote{It could be possible to constrain 
the size of such an
anomalous operator by studying the $W^*$ mode of single top
production, but
employing other search strategies than those
used in \cite{schan1}.}.  Thus, the $s$-channel single top
production process is insensitive to a modification of the
top's couplings in the form of a FCNC.

One could also imagine new physics which directly modifies the
$W$-$t$-$b$ vertex.  In this case, all three modes of single top
production will show effects from the modification, as will the
top's total decay width $\Gamma(t \ra X)$, though the branching
ratio $BR(t \ra b W^+)$ will remain close to the SM value of $\sim
100\%$.  Thus, it would again be possible to use the $W$-gluon
fusion mode to directly probe the top's partial width
$\Gamma(t \ra W^+ b)$, and from Eq. (\ref{widtheq}), compute
the full width of the top quark.
In this case, one could examine the $W^*$ production, and
$W$-gluon fusion rates separately in order to probe the dependence
of the modified vertex on $Q^2$, the momentum of the $W$ boson,
in the time-like and space-like regions, respectively.
With enough statistics, the $t W^-$ production mode could also
be helpful in this regard, since this process is not sensitive
to additional resonances, nor to FCNC interactions.

\section{$V_{tb}$ and the Possibility of New Physics}
\label{vtb}
\indent

As we have seen, it is possible that new physics effects in
the top quark sector can have a large effect on the production
of single top quarks in the $W^*$ mode (through an additional
charged heavy resonance) and in the $W$-gluon fusion mode
(through modifications of the top quark's couplings to the
partons within the proton).  This naturally leads us to
consider how one can measure the CKM parameter $V_{tb}$ with
confidence.  That is, how one can be sure that a given
measurement is accurately extracting $|V_{tb}|$, and not being
mislead by a new physics effect manifest in single top production.
               
Since both modes of single top production are sensitive to
new physics effects, we conclude that neither one alone is
enough to provide a confident extraction of $V_{tb}$.
Thus, we propose the following ratio,
\beq
R = \frac{ \sigma_{Wg} }{ \sigma_{W^*} } ,
\eeq
of the rates for the two sub-processes.  Since both cross
sections are proportional to ${|V_{tb}|}^2$, this quantity
does not depend on $V_{tb}$.  Further, it is sensitive to
a new physics effect in either cross section, and as we have
shown above, each cross section is independently sensitive to
different sources of new physics effects.  Thus this quantity
provides one with a cross-check on the confidence with which
one may regard an extraction of $V_{tb}$ from single top
production.
Assuming a top mass of
$m_t = 175$ GeV, a deviation of $R$ from its SM prediction of
2.79 by more than $\pm 25\%$ (for 2 \fb), $\pm 20\%$
(for 10 \fb), or $\pm 19\%$ (for 30 \fb) would indicate the
possible presence of new physics\footnote{The fractional
uncertainty
in $R$ is straight-forward to calculate from the uncertainties
in $\sigma_{W^*}$ and $\sigma_{Wg}$, added in quadrature.}
in single top production, and thus
call into question the validity of using it to
extract $V_{tb}$.  In that case, one can look at the two
modes for deviation from SM values in order to identify
the likely form of the new physics effect (i.e. either a heavy
charged resonance, or a modification of the top's couplings).

If new physics modifies the $W$-$t$-$b$ vertex itself, it will
affect both of the cross sections, and could cause $R$ to
remain near its SM prediction.  However, this will only be
the case if the modified $W$-$t$-$b$ vertex continues to be
a trivial function of $Q^2$, the momentum of the $W$ boson.
Since the two modes of single top production are sensitive to
different regions of $Q^2$, any modification of the
$W$-$t$-$b$ vertex with a non-trivial 
$Q^2$ dependence will likely still
cause a deviation in $R$ from its SM prediction.  In this case,
detailed study of the kinematics of the $W^*$ and
$W$-gluon fusion modes could serve to identify the $Q^2$
dependence of the modified $W$-$t$-$b$ coupling in the
time-like and space-like regions, respectively.

\begin{figure}
\centerline{\hbox{
\psfig{figure=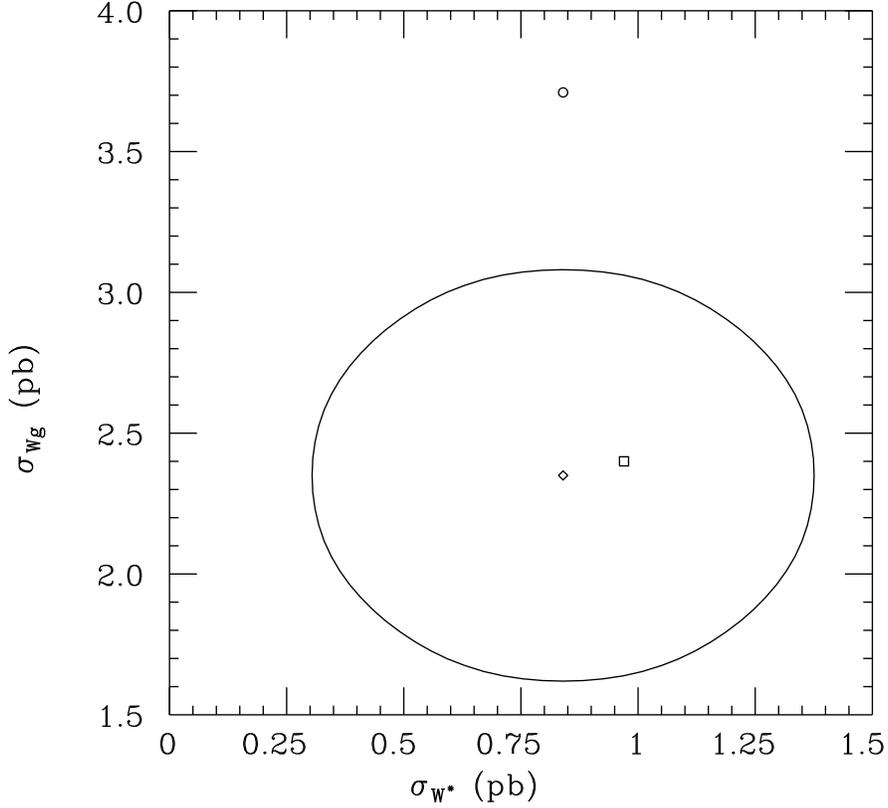,height=4.0in}}}
\caption{The location of the $m_t = 175$ GeV SM point 
(the diamond) in the $\sigma_{W^*}$-$\sigma_{Wg}$ plane,
as well as the curve representing a $3 \sigma$ deviation
from this point.  Also plotted are the results for the
top-flavor model (described in Section 3)
with $M_{Z'} = 1.2$ TeV and $\sin^2 \phi = 0.05$
(the square) and the effective Lagrangian containing
the FCNC $Z$-$t$-$c$ vertex (described in Section 4) with
$\kappa^Z_{tc} = 0.29$ (the circle).}
\label{sigmapl}
\end{figure}

It may also be useful to consider the correlation of the
two cross sections, $\sigma_{W^*}$ and $\sigma_{Wg}$,
in a two dimensional plane.
From Tables \ref{schantab} and \ref{tchantab} one can
locate the point for the SM in the $\sigma_{W^*}$ - $\sigma_{Wg}$
plane.  For example, assuming a top mass of $m_t = 175$ GeV, the
location of the SM point is at ($\sigma_{W^*}$, $\sigma_{Wg}$) = 
(0.84, 2.35).  The curves corresponding to a constant probability
for deviation from the SM values will be ellipses in this plane,
as shown in Figure \ref{sigmapl}.  From looking at this plane,
and using what we have learned about the sensitivity of the two
cross sections to new physics effects, one can also decide on a likely
cause for a given deviation.  A deviation more along the $x$-axis
is due to a larger shift in $\sigma_{W^*}$, and thus as we have
seen is more likely to be due to an additional heavy charged
resonance, while a deviation more along the $y$-axis comes
from a shift in $\sigma_{Wg}$ and thus is more likely to result
from a modification of the top quark's couplings.  The ratio
$R$ considered above is equivalent to considering a line with
unit slope running through the SM point in this plane.  By comparing
an experimental measurement of the point in this plane favored by
the data with the predictions of specific models, it could also
be possible to constrain those models.

\section{Conclusions}
\label{conclu}
\indent

In this article we have investigated the inclusive
production of single top quarks at the Tevatron Run II,
in both the
$W^*$ and $W$-gluon fusion processes, including
the first analysis of the correlated effects of
uncertainties due to top mass uncertainty, scale
dependence, and PDF choice on the uncertainties of
the cross sections.  We have also examined the
sensitivity of these processes to various types of
new physics effects.  Because the two modes of production
are sensitive to different regions of $Q^2$, we find that
they are sensitive to different types of new physics effects.
The $s$-channel $W^*$ mode is
more sensitive to an additional heavy
charged resonance, while the $W$-gluon fusion mode,
which provides a measure of the top quark's decay
width, is
more sensitive to modifications of the top quark's
couplings to the other SM particles.  As examples of
these kinds of new physics, we have considered the effect
of top-flavor models with additional
electro-weak gauge bosons, and
an anomalous $Z$-$t$-$c$ coupling to the rate of single top
production through both modes.

The possibility of new physics in the top quark sector makes
it difficult to use single top production to directly measure
the CKM parameter $V_{tb}$.  However, because the two modes are
sensitive to different forms of new physics, one can obtain
a measure of the confidence with which one can regard an
extraction of $V_{tb}$ from the single top rate by considering
their ratio, $R = \sigma_{Wg} / \sigma_{W^*}$.
A large deviation of this ratio from its SM
prediction of about 2.79 can signal the presence of new physics
in the top sector, and implies that extracting $V_{tb}$ from
single top production requires careful study to be sure that
what is being measured is actually $V_{tb}$, and not a new
physics effect.  In analyzing single top data, because of the
different sensitivities to different types of 
new physics effects of the two production modes, it is
useful to consider the experimental data in the
$\sigma_{W^*}$-$\sigma_{Wg}$ plane.  Comparison of the
predictions of explicit models with the experimental point
on this plane could be used to rule out or constrain these
models.  We have also seen that the $t W^-$ mode of single
top production is insensitive to the types of new physics
effects we have considered, and thus could provide a safe
way to measure $V_{tb}$ provided enough statistics or
a carefully tuned search strategy compensate for its low
cross section.

Thus we conclude that it is important to study single top
production at the Tevatron Run II, in both the $W^*$ and
$W$-gluon fusion modes separately, as these two modes
provide complimentary information about the top quark.
Single top production provides an excellent opportunity
to directly measure
$V_{tb}$, and to search for possible signs of the new physics
associated with the top quark.  

For completeness, we present the results for single top production
in the $W$-gluon fusion and $W^*$ modes at the LHC, using an
analysis identical to that performed for the two modes at the
Tevatron in Section \ref{singlet}.  Our results are presented
in Tables \ref{LHCs} and \ref{LHCt}.  We also note that the
$t W^-$ mode will also present another way of examining the
single top rate at the LHC, and due to the fact that it shows
different sensitivities to the types of new physics we have
considered, is also complimentary to the other two modes in
probing the properties of the top quark.

\begin{table}[htbp] 
\begin{center} 
\vskip -0.06in
\begin{tabular}{|l||c|c|c|c|}\hline \hline 
$m_{t}$ (GeV) & $\sigma_{W^*}$ (mean) (pb) & 
$\sigma_{W^*}$ (central) (pb) &
$\sigma_{W^*}$ (upper) (pb) & $\sigma_{W^*}$ (lower) (pb) \\ \hline
170 & 12.2 & 12.2 & 12.6 & 11.7 \\
171 & 11.9 & 11.9 & 12.3 & 11.4 \\     
172 & 11.7 & 11.6 & 12.0 & 11.2 \\ 
173 & 11.4 & 11.3 & 11.7 & 10.9 \\ 
174 & 11.2 & 11.1 & 11.5 & 10.7 \\ 
175 & 11.0 & 10.9 & 11.3 & 10.5 \\
176 & 10.7 & 10.7 & 11.0 & 10.3 \\ 
177 & 10.5 & 10.5 & 10.8 & 10.1 \\ 
178 & 10.3 & 10.3 & 10.6 & 9.9  \\ 
179 & 10.1 & 10.1 & 10.4 & 9.7  \\ 
180 & 9.9  & 9.9  & 10.2 & 9.5  \\ 
181 & 9.7  & 9.7  & 9.9  & 9.3  \\ 
182 & 9.5  & 9.4  & 9.7  & 9.1  \\ \hline \hline 
\end{tabular} 
\end{center} 
\vskip 0.08in 
\caption{The central and mean values
of $\sigma_{W^*}$ (in pb) at the LHC,
along with
the upper and lower bounds derived by varying the scale
and by considering the PDF sets CTEQ4M and MRRS(R1).
The central value is the value midway between the upper and lower
bounds, while the mean is the averaged result of $\sigma_{W^*}$
calculated using the CTEQ4M and MRRS(R1) PDF with the
canonical scale choice discussed in the text.}
\label{LHCs}
\end{table}

\begin{table}[htbp] 
\begin{center} 
\vskip -0.06in
\begin{tabular}{|l||c|c|c|c|}\hline \hline 
$m_{t}$ (GeV) & $\sigma_{Wg}$ (mean) (pb) & 
$\sigma_{Wg}$ (central) (pb) &
$\sigma_{Wg}$ (upper) (pb) & $\sigma_{Wg}$ (lower) (pb) \\ \hline
170 & 253 & 249 & 261 & 236 \\
171 & 250 & 248 & 259 & 234 \\     
172 & 246 & 247 & 257 & 233 \\ 
173 & 244 & 246 & 255 & 231 \\ 
174 & 241 & 244 & 253 & 230 \\ 
175 & 239 & 243 & 251 & 228 \\
176 & 237 & 241 & 249 & 226 \\ 
177 & 236 & 239 & 247 & 225 \\ 
178 & 234 & 237 & 246 & 224 \\ 
179 & 234 & 235 & 244 & 222 \\ 
180 & 233 & 232 & 242 & 221 \\ 
181 & 233 & 229 & 241 & 219 \\ 
182 & 232 & 226 & 239 & 218 \\ \hline \hline 
\end{tabular} 
\end{center} 
\vskip 0.08in 
\caption{The central and mean values
of $\sigma_{Wg}$ (in pb) at the LHC,
along with
the upper and lower bounds derived by varying the scale
and by considering the PDF sets CTEQ4M and MRRS(R1).
The central value is the value midway between the upper and lower
bounds, while the mean is the averaged result of $\sigma_{W^*}$
calculated using the CTEQ4M and MRRS(R1) PDF with the
canonical scale choice discussed in the text.}
\label{LHCt}
\end{table}

\section*{Acknowledgments }
\indent \indent
The authors would like to thank C. Balazs for helpful
discussion.
T. Tait is grateful for conversations with E. L. Berger,
S. Mrenna, and S. Murgia.  C.--P. Yuan wishes to
thank the CTEQ collaboration for useful discussions.
This work  was supported in part by the NSF grant
No. PHY-9507683.    
Part of T. Tait's work was completed at
Argonne National Laboratory,
in the High Energy Physics division and was supported in
part by the U.S. Department of Energy, High Energy Physics
Division, under Contract W-31-109-Eng-38.


\begin{thebibliography}{99}
\frenchspacing

\bibitem{topdisc}
F.~Abe {\it et al.}, {Phys. Rev. Lett.} {\bf 73}, 225 (1994); \\
S.~Abachi {\it et al.},{Phys. Rev. Lett.} {\bf 72}, 2138 (1994);\\
CDF Collaboration, {Phys. Rev. Lett.} {\bf 74}, 2626 (1995);\\
D0 Collaboration, {Phys. Rev. Lett.}  {\bf74}, 2632 (1995).

\bibitem{ttbar}
P. Nason, S. Dawson, and R. K. Ellis, Nucl. Phys.
{\bf B303}, 607 (1988);
Nucl. Phys. {\bf B327}, 49 (1989);
Nucl. Phys. {\bf B335}, 260 (1990).\\
G. Altarelli, M. Diemoz, G. Martinelli, and P. Nason,
Nucl. Phys. {\bf B308}, 724 (1988).\\ 
W. Beenakker, H. Kuijf, W.L. van Neerven, and J. Smith, Phys. Rev.
{\bf D 40}, 54 (1989).\\
W. Beenakker, W.L. van Neerven, R. Meng, 
G.A. Schuler, and J. Smith, Nucl. Phys.
{\bf B351}, 507 (1991).\\
R.K. Ellis, Phys. Lett. {\bf B259}, 492 (1992).\\
E. Laenen, J. Smith, and W.L. van Neervan, Nucl. Phys.
{\bf B369}, 543 (1992);
Phys. Lett. {\bf B321}, 254 (1994).\\
N. Kidonakis and J. Smith, Phys. Rev. {\bf D 51}, 6092 (1995).
S. Catani, M. Mangano, P. Nason, and L. Trentadue, Phys. Lett.
{\bf B378}, 329 (1996); Nucl. Phys. {\bf B478}, 273 (1996).\\
E. Berger and H. Contopanago, Phys. Lett. {\bf B 361},
115 (1995); Phys. Rev. {\bf D54}, 3085 (1996); hep-ph/9706206.

\bibitem{schan}
R. Cortese and R. Pertronzio, Phys. Lett. {\bf B253}, 494 (1991).

\bibitem{schan1}
T. Stelzer and S. Willenbrock, Phys. Lett. {\bf B357}, 125 (1995).

\bibitem{schan2}
M. C. Smith and S. Willenbrock, Phys. Rev. {\bf D 54}, 6696 (1996).

\bibitem{schanresum}
S. Mrenna and C.--P. Yuan, "Effects of QCD Resummation on $W^+h$
and $t \bar b$ Production at the Tevatron", hep-ph/9703224 (1997).

\bibitem{boos}
A. P. Heinson, A. S. Belyaev, and E. E. Boos, Phys. Rev.
{\bf D 56}, 3114 (1997).

\bibitem{tchan}
S. Dawson, Nucl. Phys. {\bf B249}, 42 (1985).\\
S. Willenbrock and D. Dicus, Phys. Rev. {\bf D 34}, 155 (1986).

\bibitem{tchan1}
C.--P. Yuan, Phys. Rev. {\bf D 41}, 42 (1990).
   
\bibitem{tchan2}
R. K. Ellis and S. Parke, Phys. Rev. {\bf D 46}, 3785 (1992).

\bibitem{cpy} 
C.--P. Yuan, "Top Quark Physics at Hadron Colliders", published in
CCAST Symposium 1993, 259 (1993); hep-ph/9308240.\\
D. O. Carlson and C.--P. Yuan, "Probing New Physics from the Single
Top Production", Particle Phys. \& Phen. 1995, 172 (1995);
hep-ph/9509208.\\
C.--P. Yuan, "Top Quark Physics", published in Valencia Elem. Part.
Phys. 1995, 148 (1995); hep-ph/9509209.\\
C.--P. Yuan, "Physics of Top Quark at the Tevatron", talk
given at 5th Mexican Workshop of Particles and Fields, Puebla,
Mexico, Oct. 30 - Nov 3, 1995; hep-ph/9604434.\\

\bibitem{dougthesis}
D. Carlson, Ph.D. thesis, Michigan State University, MSUHEP-050727,
August 1995.

\bibitem{bordes}
G. Bordes and B. van Eijk, Z. Phys. {\bf C 57}, 81 (1993).\\
G. Bordes and B. van Eijk, Nucl. Phys. {\bf B435}, 23 (1995).

\bibitem{tchan3}
T. Stelzer, Z. Sullivan, and S. Willenbrock, "Single-Top
Production via $W$-Gluon Fusion at Next-to-Leading Order",
hep-ph/9705398 (1997).

\bibitem{tw}
G. Ladinsky and C.--P. Yuan, Phys. Rev. {\bf D 43}, 789 (1991).\\
S. Moretti, Cavendish-HEP-97/05 (1997); hep-ph/9705388.

\bibitem{cteq4}
CTEQ Collaboration: H. Lai, J. Huston, S. Kuhlmann,
F. Olness, J. Owens, D. Sopher, W.-K. Tung, and H. Weerts,
Phys. Rev. {\bf D 55}, 1280 (1997).

\bibitem{mrsa}
A. Martin, R. Roberts, M. G. Ryskin,
and W. J. Stirling, "Consistent Treatment of Charm Evolution
in Deep Inelastic Scattering", hep-ph/9612449.

\bibitem{effW}
R.N. Cahn and S. Dawson, {\it Phys. Lett.}
{\bf B136}, 196 (1984), {\bf B138}, 464(E) (1984); \\
M.S. Chanowitz and M.K. Gaillard, {\it Phys. Lett.} {\bf B142},
85 (1984);\\
G.L. Kane, W.W. Repko and W.R. Rolnick, {\it Phys. Lett.}
 {\bf B148}, 367 (1984); \\
S. Dawson, {\it Nucl. Phys.} {\bf B249}, 427 (1985);\\
J. Lindfors, {\it Z. Phys.} {\bf C28}, 427 (1985);\\
W.B. Rolnick, {\it Nucl. Phys.} {\bf B274}, 171 (1986);\\
P.W. Johnson, F.I. Olness and W.-K. Tung, {\it Phys. Rev.}
 {\bf D36}, 291 (1987);\\
Z. Kunszt and D.E. Soper, {\it Nucl. Phys.} {\bf B296}, 253 (1988);\\
A. Abbasabadi, W.W. Repko, D.A. Dicus and R. Vega,
{\it Phys. Rev.} {\bf D38}, 2770 (1988);
S. Dawson, {\it Phys. Lett.} {\bf B217}, 347 (1989); \\
 S. Cortese and R. Petronzio,  {\it Phys. Lett.} {\bf B276}, 203
(1992);\\
I. Kuss and H. Spiesberger, {\it Phys. Rev.} {\bf D53}, 6078 (1996).

\bibitem{pdb}
Particle Data Group, Phys. Rev. {\bf D 54}, 1 (1996).

\bibitem{wvtb}
S. Willenbrock, "Top Quark Physics for Beautiful and
Charming Physicists", hep-ph/9709355 (1997).

\bibitem{tev2000}
Dan Amidei and Chip Brock, ``Report of the $TeV 2000$ Study
Group on Future ElectroWeak Physics at the Tevatron'', 1995.

\bibitem{tf}
X. Li and E. Ma, Phys. Rev. Lett. {\bf 47}, 1788 (1988);
{\it ibid.} {\bf 60}, 495 (1988).\\
X. Li and E. Ma, Phys. Rev. {\bf D 46}, 1905 (1992).\\
X. Li and E. Ma, J. Phys. {\bf G19}, 1265 (1993).

\bibitem{tf1}
E. Malkawi, T. Tait, and C.--P. Yuan,
Phys. Lett. {\bf B385}, 304 (1996).

\bibitem{tf2}
D. J. Muller and S. Nandi, Phys. Lett. {\bf B383}, 345 (1996).

\bibitem{es}
E. Simmons, Phys. Rev. {\bf D 55}, 5494 (1997).

\bibitem{cmy}
D. Carlson and C.--P. Yuan, Phys. Lett. {\bf B306}, 386 (1993).\\
D. Carlson, E. Malkawi and C.--P. Yuan, Phys. Lett. {\bf B337} 145
(1994).\\
F. Larios and C.--P. Yuan, Phys. Rev. {\bf D 55}, 7218 (1997). \\
F. Larios, T. Tait, and C.--P. Yuan, hep-ph/9709316 (1997).

\bibitem{Ztc}
T. Han, R.D. Peccei, and X. Zhang,
Nucl. Phys. {\bf B454}, 527 (1995).

\bibitem{tcgam}
T. Han, K. Whisnant, B.-L. Young, X. Zhang, Phys. Rev.
{\bf D 55}, 7241 (1997).\\
K. J. Abraham, K. Whisnant, B.-L. Young, hep-ph/9707476 (1997).

\bibitem{tcg}
E. Malkawi and T. Tait, Phys. Rev. {\bf D 54}, 5758 (1996).\\
T. Han, K. Whisnant, B.-L. Young, X. Zhang, Phys. Lett.
{\bf B 385}, 311 (1996).\\
T. Tait and C.-P. Yuan, Phys. Rev. {\bf D 55}, 7300 (1997).

\bibitem{Zbb}
A. Datta and X. Zhang, Phys. Rev. {\bf D 55}, 2530 (1997).
        
\bibitem{nonlinear}
R. D. Peccei and X. Zhang, Nucl. Phys. {\bf B337}, 269 (1990).

\end{thebibliography}
\end{document}